\documentclass[11pt]{article}
\usepackage{graphicx}

\newsavebox{\hflrar}
\sbox{\hflrar}{\makebox[0pt][l]
{${\scriptstyle \leftharpoonup}$}{${\scriptstyle \rightharpoonup}$}}

\def \to {\rightarrow}

\def\bfsig{\mbox{\boldmath$\sigma$}}

\begin{document}
\begin{flushright}
AS-ITP-2003-010
\end{flushright}
\pagestyle{plain}
\vskip 10mm
\begin{center}
{\bf\Large Factorization Approach for Inclusive Production of
Doubly Heavy Baryon } \\
\vskip 10mm
J.P. Ma   \\
{\small {\it Institute of Theoretical Physics , Academia
Sinica, Beijing 100080, China }} \\
~~~ \\
Z.G. Si \\
{\small {\it Department of Physics, Shandong University, Jinan Shandong 250100, China}}
\end{center}
\vskip 0.4 cm
%%%%%%%%%%%%%%%%%% abstract of this paper %%%%%%%%%%%%%%%%%%%%%%%%%%%%%%%%

\begin{abstract}
We study inclusive production of doubly heavy baryon at
a $e^+e^-$ collider and at hadron colliders through fragmentation.
We study the production by factorizing nonpertubative- and
perturbative effects.
In our approach the production can be thought
as a two-step process:  A pair of heavy quarks
can be produced perturbatively and then the pair is transformed into the
baryon. The transformation is nonperturbative. Since a heavy quark moves with a small
velocity in the baryon in its rest frame, we can use NRQCD to describe
the transformation and perform a systematic expansion in the small velocity.
At the leading order we find that the baryon
can be formed from two states of the heavy-quark pair,
one state is with the pair in $^3S_1$ state and in color ${\bf \bar 3}$,
another is with the pair in $^1S_0$ state and in color ${\bf 6}$.
Two matrix elements are defined for the
transformation from the two states, their perturbative coefficients
in the contribution to the cross-section at a $e^+e^-$ collider
and to the function of heavy quark fragmentation are calculated.
Our approach is different than previous approaches where
only the pair in $^3S_1$ state and in color ${\bf \bar 3}$ is taken into account.
Numerical results for $e^+e^-$ colliders at the two $B$-factories
and for hadronic colliders LHC and Tevatron are given.
\vskip 5mm \noindent
%PACS numbers: 12.38.Bx, 12.39.Hg, 13.85.Ni
\par\noindent
%Key Words: Leading Particle Effect, Quark Recombination, Twist 4 and
% Heavy Quark Effective Theory.
\end{abstract}
\par\vfil\eject
\noindent
{\bf\large 1. Introduction}
\par\vskip20pt
It is a well known fact that the structure of a heavy hadron containing
one- or more heavy quarks is much simpler than that of light hadrons,
hence, theoretical study of a heavy hadron can be done more rigorously than
that of a light hadron. In the last decade, the heavy quark effective theory
was derived from QCD\cite{HQET}, and it was widely used for hadrons containing
one heavy quark $Q$. For hadrons containing a heavy quark $Q$ and a heavy
antiquark $\bar Q$, i.e., quarkonia, nonrelativistic QCD(NRQCD)
provided a systematical, model-independent way to study them\cite{nrqcd,nrqcd1}.
The existence of heavy hadrons containing one heavy quark and quarkonia
is well confirmed in experiment, while the existence of heavy baryons
containing two heavy quarks $Q$ is not completely confirmed yet, only
one evidence for $\Xi^+_{cc}$ is found by SELEX Collaboration\cite{exp},
and it is also pointed out that the evidence may lack sufficient support\cite{comm}.
In this work we study inclusive production of a heavy baryon containing
two heavy quarks, or doubly heavy baryon, at a $e^+e^-$ collider like
BaBar and Belle, and the production through fragmentation of a heavy quark $Q$,
where a factorization of nonperturbative effects is performed.
\par
We denote $H_{QQ}$ for a heavy baryon containing two heavy quark $Q$. In the
rest frame of $H_{QQ}$ the heavy quarks move with a small velocities $v_Q$, this
enables to use NRQCD to describe heavy quarks in $H_{QQ}$ and
the nonperturbative effect related to $H_{QQ}$, where a systematic expansion in
$v_Q$ can be performed. On the other hand, the production of a
heavy quark pair $QQ$ can be studied with perturbative QCD because
the large mass $m_Q$ of the heavy quark $Q$. After its production the
pair will combine other light dynamical freedoms of QCD to form
the baryon $H_{QQ}$. In the formation the baryon $H_{QQ}$ will carry
the most momentum of the pair $QQ$, the residual momentum and that
of light dynamical freedoms are at order of $\Lambda_{QCD}$.
The above discussion indicates that
an inclusive production rate of $H_{QQ}$ can be factorized, where it consists
of two parts, one part is for production of a $QQ$ pair, determined by
perturbative QCD, another part is for nonperturbative transition
of the $QQ$ pair into $H_{QQ}$ and can be defined in terms
of NRQCD matrix elements. At leading order of $v_Q$, we find there are
two NRQCD matrix elements for the nonperturbative transition, one is
for the transition of a $QQ$ pair in $^3S_1$ state and in the color representation
${\bf \bar 3}$, another is for the transition of a $QQ$ pair in $^1S_0$ state
and in the color representation ${\bf 6}$. A power counting in $v_Q$ for these
matrix elements is made and it indicates that they are at the same order of $v_Q$.
If one takes $H_{QQ}$ as a bound state of $QQq$ only, then the transition
of a $QQ$ pair in $^1S_0$ state
and in the color representation ${\bf 6}$ is suppressed. However,
in the power counting one should pay attention to that $H_{QQ}$ is not only
as a bound state of $QQq$, but it can also be a bound state of $QQqg$, these
states are possible components of $H_{QQ}$. Unlike the case with quarkonia,
where the probability to find the component of $Q\bar Q g$ in a quarkonium
is always suppressed with the power counting rule in \cite{nrqcd1, nrqcd},
because the gluon is emitted by a heavy quark with a probability
proportional to $v_Q$, for the case with $H_{QQ}$, the probability to find
the component $QQqg$ is at the same order as that to find the component
$QQq$, because the gluon can be emitted by the light quark $q$ easily.
Therefore, the transitions of the two states of the $QQ$ pair are at the
same order of $v_Q$.
\par
Inclusive production of a doubly heavy baryon has been studied before
\cite{FLSW,K1,K2,K3,SPB,BLO}. In \cite{FLSW} the fragmentation function
of a heavy quark into $H_{QQ}$ is calculated, in \cite{K1} inclusive production
of $H_{QQ}$ at $e^+e^-$ colliders is studied by using quark-hadron duality,
in \cite{K2,K3,SPB} inclusive production at various colliders are studied.
In all these studies one always assumes that a $QQ$ pair in $^3S_1$ state
and in the color representation ${\bf\bar  3}$ is produced first and then
this pair is transformed into $H_{QQ}$. From our above discussion, one should
also add the contribution from the contribution of the transition of
a $QQ$ pair in $^1S_0$ state
and in the color representation ${\bf 6}$. In previous studies a wave function
for the $QQ$ pair in $^3S_1$ state with color ${\bf\bar 3}$ is introduced
to characterize the nonperturbative transition. In this work we characterize
nonperturbative transitions with NRQCD matrix elements, which are well defined
and can be conveniently studied with nonperturbative methods like
QCD sum rule method. Based our new results we give a prediction for
production rate at $e^+e^-$ colliders at the two $B$-factories, and
at hadron colliders like Tevatron and LHC.
\par
Our work is organized as the following: In Sect. 2. we study inclusive production
at a $e^+e^-$ collider, where we perform the mentioned factorization and
find out two NRQCD matrix elements for nonperturbative effects at leading order
of $v_Q$. A discussion about power counting in $v_Q$ for these matrix elements
is given. Numerical predictions for $\Xi^+_{cc}$ are presented. In Sect. 3. we calculate
the fragmentation function of $Q$ into $H_{QQ}$ by starting from definition
of fragmentation functions and give an estimation for production rate
at hadron colliders for $\Xi^+_{cc}$ with large transverse momentum. Sect.4.
is our summary.
\par

\par\vskip20pt
\noindent
{\bf\large 2. Production in $e^+e^-$ collision}
\par\vskip20pt
We consider the process:
\begin{equation}
e^+(p_1) + e^- (p_2) \to \gamma^* (q) \to  H_{QQ}(k) +X,
\end{equation}
where the heavy baryon $H_{QQ}$ contains two  heavy quarks $Q$.
We can always divide the unobserved state
into a nonperturbatively produced part $X_N$ and a perturbatively produced
part $X_P$, i.e., $X=X_N+X_P$. At tree level, the perturbatively produced
part $X_P$ consists of two heavy antiquark $\bar Q$,
the scattering amplitude for the process can be written:
\begin{equation}
{\cal T} = \frac{1}{2}\int \frac{d^4k_1}{(2\pi)^4}
        A_{ij}(k_1, k_2,p_3,p_4)
              \int d^4x_1 e^{-ik_1\cdot x_1}
              \langle H_{QQ}(k) +X_N\vert
               \bar Q_i(x_1) \bar Q_j(0) \vert 0\rangle,
\end{equation}
where indices $i,j$ are Dirac- and color indices, $Q(x)$ is the Dirac field
for the heavy quark $Q$. The perturbative amplitude is given by diagrams
in Fig.1.
If one replaces in Eq.(2) the state $\langle H_{QQ}(k) +X_N\vert$ with
a state of two free quark $Q$ with the momentum $k_1$ and $k_2$
respectively, one will obtain the amplitude ${\cal T}$ as
the amplitude for $e^+(p_1) + e^- (p_2) \to \gamma^* (q) \to  Q(k_1)+Q(k_2)
+\bar Q(k_3) +\bar Q(k_4)$.
\par
%%%%%%%%%%%%%%%% Inset Fig. 1 here %%%%%%%%%%%%%%%%%%%%%%%%%%%%%%

\begin{figure}[hbt]
\centering
\includegraphics[width=8cm]{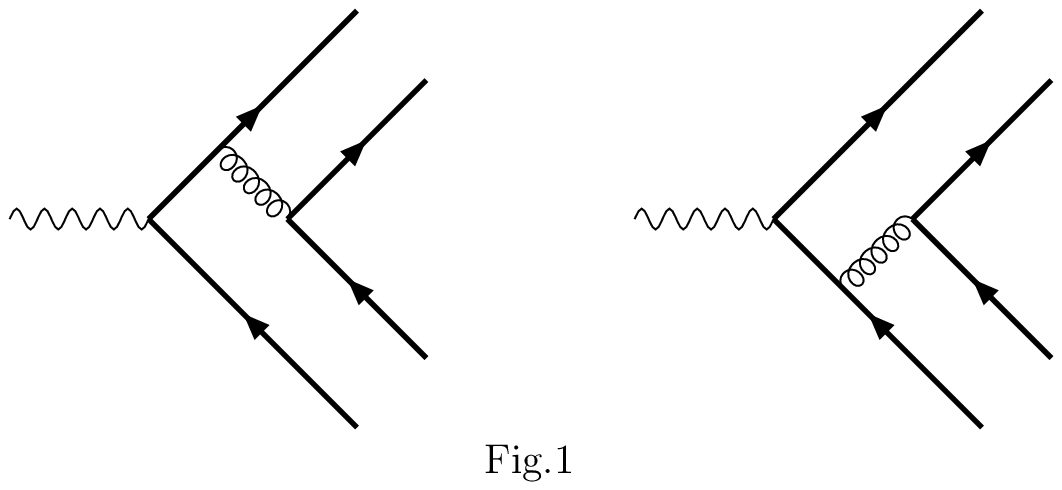}
\caption{Feynman diagrams for the amplitude $A$, other two diagrams
are obtained by exchange the momenta of antiquarks.}
\label{Feynman-dg1}
\end{figure}
\par
%%%%%%%%%%%%%%%%%%%%%%%%%%%%%%%%%%%%%%%%%%%%%%%%%%%%%%%%%%%%%%%%%%%%
\par
With the amplitude the differential cross-section
for the process in Eq.(1) can be written as:
\begin{eqnarray}
 d\sigma &=&\frac{1}{2}\frac{1}{2\hat s} \sum_{X_N}
    \frac {d^3k}{(2\pi)^3}\int \frac{d^3p_3}{(2\pi)^3}\frac{d^3p_4}{(2\pi)^3}
       (2\pi)^4\delta^4(p_1+p_2-k-p_3-p_4-P_{X_N})\nonumber\\
      &&\cdot  \frac{1}{4}\int \frac{d^4k_1}{(2\pi)^4} \frac{d^4k_3}{(2\pi)^4}
         A_{ij}(k_1,k_2,p_3,p_4) (\gamma^0 A^\dagger
         (k_3,k_4,p_3,p_4)\gamma^0)_{kl} \nonumber\\
      &&\cdot \int d^4x_1 d^4x_3 e^{-ik_1\cdot x_1+ik_3\cdot x_3}
      \langle 0\vert \bar Q_k(0) Q_l(x_3) \vert H_Q+X_N\rangle
      \langle H_Q +X_N\vert \bar Q_i(x_1) \bar Q_j(0) \vert 0\rangle,
\end{eqnarray}
where
the average over spin of initial leptons and summation over the spin of the
baryon $H_{QQ}$
and over color-, spin state of two $\bar Q$ quarks
is implied. The factor $1/2$ is because of two identical antiquark.
In this section we take nonrelativistic normalization
for heavy quarks and the heavy baryon.
Using translational covariance one can eliminate the sum over $X_N$.
We define $a^\dagger ({\bf k})$ as the creation operator for $H_{QQ}$
with the three momentum ${\bf k}$ and we
obtain:
\begin{eqnarray}
 d\sigma &=& \frac{1}{2\hat s}\frac {d^3k}{(2\pi)^3}
  \int \frac{d^3p_3}{(2\pi)^3}\frac{d^3p_4}{(2\pi)^3}
      \cdot  \int \frac{d^4k_1}{(2\pi)^4} \frac{d^4k_3}{(2\pi)^4}
         A_{ij}(k_1,k_2,p_3,p_4) (\gamma^0 A^\dagger
         (k_3,k_4,p_3,p_4)\gamma^0)_{kl} \nonumber\\
      &&\cdot \frac{1}{8}\int d^4x_1 d^4x_2 d^4x_3 e^{-ik_1\cdot x_1-ik_2\cdot x_2
       +ix_3\cdot k_3}
      \langle 0\vert  Q_k(0) Q_l(x_3)  a^\dagger ({\bf k})
       a({\bf k}) \bar Q_i(x_1) \bar Q_j(x_2) \vert 0\rangle.
\end{eqnarray}
This contribution can be represented by Fig.2, where
the black box represents the Fourier transformed matrix element.
\par
%%%%%%%%%%%%%%%% Inset Fig. 2 here %%%%%%%%%%%%%%%%%%%%%%%%%%%%%%
\begin{figure}[hbt]
\centering
\includegraphics[width=7cm]{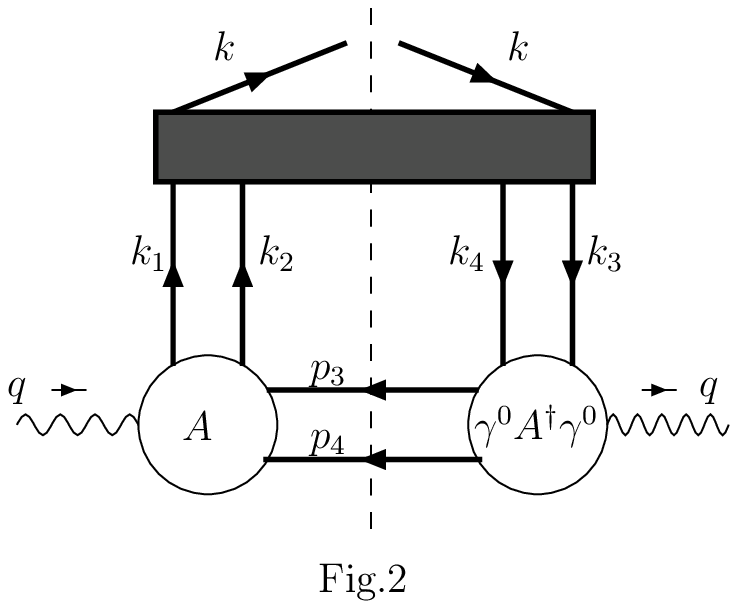}
\caption{Graphic representation for the contribution in Eq.(4), the broken
line is the cut and $k_4=k_1+k_2-k_3$.
}
\label{Feynman-dg2}
\end{figure}
\par
%%%%%%%%%%%%%%%%%%%%%%%%%%%%%%%%%%%%%%%%%%%%%%%%%%%%%%%%%%%%%%%%%%%%
Since heavy quarks move with a small velocity $v_Q$ inside of the baryon
in its rest frame, one can use NRQCD to handel heavy quarks, in which
a systematic expansion in $v_Q$ can be used. Hence,
the Fourier transformed matrix element can be expanded in $v_Q$ with fields
of NRQCD. The relation between NRQCD fields and Dirac field $Q(x)$
in the baryon's rest frame is
\begin{equation}
Q(x)= e^{-im_Q t} \left\{
\begin{array}{ll}
 \psi (x) \\
\ 0
\end{array}
\right\} +{\cal O}(v_Q) +\cdots,
\end{equation}
where $\cdots$ denote the part for antiquark, which is irrelevant here.
We will work at the leading order of $v_Q$. To express our results for
the Fourier transformed matrix element in a covariant way, we denote
$v$ as the velocity of the baryon with $v^\mu =k^\mu/M_{H_{QQ}}$.
The Fourier transformed matrix element is then related to that in the
rest frame:
\begin{eqnarray}
 && v^0 \int d^4x_1 d^4x_2 d^4x_3 e^{-ik_1\cdot x_1-ik_2\cdot x_2
       +ix_3\cdot k_3}
      \langle 0\vert  Q_k(0) Q_l(x_3)  a^\dagger ({\bf k})
       a({\bf k}) \bar Q_i(x_1) \bar Q_j(x_2) \vert 0\rangle
 \nonumber\\
  && = \int d^4x_1 d^4x_2 d^4x_3 e^{-ik_1\cdot x_1-ik_2\cdot x_2
       +ix_3\cdot k_3}
      \langle 0\vert  Q_k(0) Q_l(x_3)  a^\dagger ({\bf k}=0)
       a({\bf k}=0) \bar Q_i(x_1) \bar Q_j(x_2) \vert 0\rangle.
\end{eqnarray}
Using the expansion in Eq.(5) for the matrix element, one will obtain
the matrix element containing NRQCD fields $\psi (x)$ and $\psi^\dagger (x)$,
the space-time of the matrix element with NRQCD fields is controlled by the
scale $m_Q v_Q$ or $\Lambda_{QCD}$, hence at leading order of $v_Q$ one can
neglect the space-time dependence, also at this order the baryon mass
$M_{H_{QQ}}$ is approximated by $2m_Q$. With the approximation
the matrix element in Eq.(5)
is related to the matrix element of NRQCD:
\begin{equation}
\langle 0\vert  \psi^{a_3}_{\lambda_3}(0) \psi^{a_4}_{\lambda_4}(0)  a^\dagger
       a \psi^{a_1\dagger}_{\lambda_1}(0) \psi^{a_2\dagger}_{\lambda_2}(0) \vert 0\rangle,
\end{equation}
where we suppressed the notation ${\bf k}=0$ and it is always implied
that NRQCD matrix elements are defined in the rest frame of $H_{QQ}$.
In the above equation we use $a_{i}(i=1,2,3,4)$ to label the color
of quarks fields, while $\lambda_i (i=1,2,3,4)$ is for spin indices.
The spin indices of the above matrix element runs from $1$ to $2$
because of the structure in Eq.(5). By using rotation invariance,
color-symmetry and Pauli principle of two identical fermions
the above matrix element is parameterized by two parameters:
\begin{eqnarray}
\langle 0\vert  \psi^{a_3}_{\lambda_3} \psi^{a_4}_{\lambda_4}  (a^\dagger
a ) \psi^{a_1\dagger}_{\lambda_1} \psi^{a_2\dagger}_{\lambda_2} \vert 0\rangle
  &=& (\varepsilon)_{\lambda_4\lambda_3} (\varepsilon)_{\lambda_2\lambda_1}
      \cdot (\delta_{a_1a_4}\delta_{a_2a_3}+\delta_{a_1a_3}\delta_{a_2a_4})
      \cdot h_1
    \nonumber\\
  && \  +(\sigma^n\varepsilon)_{\lambda_4\lambda_3}
        (\varepsilon\sigma^n)_{\lambda_2\lambda_1}
      \cdot (\delta_{a_1a_4}\delta_{a_2a_3}-\delta_{a_1a_3}\delta_{a_2a_4})
      \cdot h_3,
\end{eqnarray}
where $\sigma^i (i=1,2,3)$ are Pauli matrices, $\varepsilon =i\sigma^2$
is totally anti-symmetric. The parameters are defined as:
\begin{eqnarray}
h_1 &=& \frac{1}{48}
   \langle 0\vert [ \psi^{a_1}\varepsilon \psi^{a_2}
                +  \psi^{a_2}\varepsilon \psi^{a_1}]  a^\dagger
a \psi^{a_2\dagger} \varepsilon \psi^{a_1\dagger} \vert 0\rangle,
\nonumber\\
h_3 &=& \frac{1}{72}
   \langle 0\vert [ \psi^{a_1}\varepsilon \sigma^n \psi^{a_2}
                -  \psi^{a_2}\varepsilon\sigma^n \psi^{a_1}]  a^\dagger
a \psi^{a_2\dagger} \sigma^n \varepsilon\psi^{a_1\dagger} \vert 0\rangle,
\end{eqnarray}
the physical interpretation of parameters is clear: $h_1$ represents
the probability for a $QQ$ pair in a $^1S_0$ state and in the color state
of ${\bf 6}$ to transform into the baryon, while $h_3$ represents
the probability for a $QQ$ pair in a $^3S_1$ state and in the color state
of ${\bf\bar 3}$ to transform into the baryon. With these results
the Fourier transformed matrix element in Eq.(6) can be expressed
as:
\begin{eqnarray}
 && v^0 \int d^4x_1 d^4x_2 d^4x_3 e^{-ik_1\cdot x_1-ik_2\cdot x_2
       +ix_3\cdot k_3}
      \langle 0\vert  Q^{a_3}_k(0) Q^{a_4}_l(x_3)  a^\dagger ({\bf k})
       a({\bf k}) \bar Q^{a_1}_i(x_1) \bar Q^{a_2}_j(x_2) \vert 0\rangle
 \nonumber\\
  && =(2\pi)^4\delta^4 (k_1-m_Qv)(2\pi)^4\delta^4 (k_2-m_Qv)(2\pi)^4\delta^4 (k_3-m_Qv)
\nonumber \\
  && \ \cdot \big[ -(\delta_{a_1a_4}\delta_{a_2a_3}+\delta_{a_1a_3}\delta_{a_2a_4})
          (\tilde P_v C\gamma_5 P_v)_{ji}(P_v \gamma_5 C \tilde P_v)_{lk}\cdot h_1
\nonumber\\
&& \  + (\delta_{a_1a_4}\delta_{a_2a_3}-\delta_{a_1a_3}\delta_{a_2a_4})
          (\tilde P_v C\gamma^\mu  P_v)_{ji}(P_v \gamma^\nu C \tilde P_v)_{lk}(
            v_\mu v_\nu-g_{\mu\nu})\cdot  h_3\big ] +\cdots,
\end{eqnarray}
with
\begin{equation}
P_v =\frac{1+\gamma\cdot v}{2}, \ \ \ \  \tilde P_v=\frac{1+\tilde\gamma\cdot v}{2}.
\end{equation}
In the above equation we used $a_{i}(i=1,2,3,4)$ for the color indices
and $ijkl$ for the Dirac indices. $\tilde A$ denotes the transpose of the matrix $A$.
In Eq.(10) $\cdots$ denote terms at higher
orders of $v_Q$, which are neglected in this work. These terms as corrections
can be systematically added. $C=i\gamma^2\gamma^0$
is the matrix for charge conjugation. Substituting Eq.(10) into Eq.(4),
one can easily find
that the two heavy quarks $Q$ are projected to on-shell states and they have
the same momentum $m_Qv$.
\par
Numerical values of the two matrix elements are unknown yet. There are attempts
to relate $h_3$ to the corresponding matrix element $\vert \langle  0\vert
\chi^\dagger\bfsig \psi \vert ^3S_1 \rangle \vert ^2$ for the transition
of a $Q\bar Q$ pair into a $^3S_1$ quarkonium, in which one introduces a wave
function for $^3S_1$ $QQ$ state, the radial wave function $R_{QQ}(r)$ at origin
is related to $h_3$ by:
\begin{equation}
h_3 =\frac{1}{4\pi} \vert R_{QQ}(0)\vert^2.
\end{equation}
Assuming the potentials for binging $Q\bar Q$ and $QQ$ state are hydrogen-like,
then the difference between the potential for $Q\bar Q$ and that for $QQ$
is determined by color structures, using the difference a relation
between $R_{QQ}(0)$ and $R_{Q\bar Q}(0)$ can be obtained. But such a relation
can not be found for $h_1$, and the potentials are not exactly hydrogen-like
because of QCD confinement. An rough estimation can be obtained by noting that
one can give a power counting in $v_Q$ for these matrix elements, similarly to
the power counting of those matrix elements for quarkonia. For this we note
that in general $H_{QQ}$ is a bound state of two heavy quarks $Q$ with other light
dynamical freedoms of QCD, the state can be written as:
\begin{equation}
\vert H_{QQ} \rangle = c_1 \vert QQ q\rangle +c_2 \vert QQ qg \rangle
   +c_3 \vert QQ q gg \rangle +\cdots.
\end{equation}
For a $QQ$ pair in $^3S_1$ state with the color ${\bf \bar 3}$,
one of the heavy quarks can emits a gluon, which does not change
the spin of the heavy quark, and this gluon then splits into a $q\bar q$.
The heavy quark pair can combine the light quark $q$ to form $H_{QQ}$, while
$\bar q$ will combine other partons to transform into unobserved states. Since
the probability of a heavy quark emitting such a gluon is proportional
to $v_Q$, then we have
\begin{equation}
 h_3 \sim v_Q^2 \vert \langle  0\vert
\chi^\dagger\bfsig \psi \vert ^3S_1 \rangle \vert ^2.
\end{equation}
For a $QQ$ pair in $^1S_0$ state with the color ${\bf 6}$, if one follows
the above discussion by requiring that $H_{QQ}$ is formed by the component
$\vert QQ q\rangle$, then the emitted gluon must change the spin of
the heavy quark, then one may conclude that
$h_1$ is at higher order than $v_Q^2$ in comparison with $h_3$. However,
$H_{QQ}$ can be formed with the component $\vert QQ qg \rangle$,
this component can be formed as the following: One of the heavy quarks emits a gluon,
which does not change
the spin of the heavy quark, and this gluon splits into a $q\bar q$,
the light quarks can also emit gluons, then the component
can be formed with the light quark $q$ plus one gluon, other light partons
are transformed into unobserved state. Unlike for quarkonium systems, in which
the leading component is the $Q\bar Q$ state, for $H_{QQ}$, all components
must contain at least one light quark $q$. Because a light quark can emit gluons
easily, the components in Eq.(13) are important at the same level, i.e.,
$c_1\sim c_2 \sim c_3 \cdots$. Hence we have:
\begin{equation}
 h_1 \sim v_Q^2 \vert \langle  0\vert
\chi^\dagger\bfsig \psi \vert ^3S_1 \rangle \vert ^2.
\end{equation}
In deriving the above power counting we basically used perturbative QCD,
in general one can not use perturbative QCD to discuss how $H_{QQ}$ is formed,
but for power counting in $v_Q$ it gives right answers.
\par
With the results in Eq.(10) we obtain the differential cross
section:
\begin{eqnarray}
d\sigma &=&\frac{1}{2\hat{s}}\frac{d^3k}{(2\pi)^3 v_0}
\int\frac{d^3p_3}{(2\pi)^32E_3} \frac{d^3p_4}{(2\pi)^3 2E_4}
(2\pi)^4\delta^4(p_1+p_2-p_3-p_4-k) \nonumber \\
&&\Big(\frac{4e^4 g_s^4 e_Q^2}{3\hat{s}^2}\Big)\frac{1}{m_Q^2}
\Big\{h_1 {\cal B}^{(1)}+16h_3 {\cal B}^{(3)}\Big\},
\end{eqnarray}
where
\begin{eqnarray}
{\cal B}^{(i)}&=&\frac{8}{(1-x_3)^2(1-x_4)^2(x_3+x_4)^2}
\Big\{{\cal A}_1^{(i)}+
\frac{8(p_1\cdot p_3)^2}{s^2} {\cal A}_2^{(i)}+
\frac{8(p_1\cdot p_4)^2}{s^2}
{\cal A}_2^{(i)}|_{x_3\leftrightarrow x_4}\nonumber\\
&&+ \frac{8(p_1\cdot p_3)(p_1\cdot p_4)}{s^2} {\cal A}_3^{(i)}
+\frac{2(p_1\cdot p_3)}{s}{\cal A}_4^{(i)}
+\frac{2(p_1\cdot p_4)}{s}{\cal A}_4^{(i)}|_{x_3\leftrightarrow x_4}
\Big\}.
\end{eqnarray}
The functions ${\cal A}^{(1,3)}_i(i=1,2,3,4)$ depend on
$x_{3,4}=2p^0_{3,4}/\sqrt{s}$ with $s=(p_1+p_2)^2$ and are
given in the appendix. Now we take the heavy quark $Q$ as
charm quark $c$ to give some numerical results for
$\Xi^+_{cc}$. It should be noted that the results also apply
for $\Xi^{++}_{cc}$ because isospin symmetry.
we obtain the total cross section
for $e^+ e^-$ colliders at the two B factories with $\sqrt{s} =10.6$ GeV:
\begin{equation}
\sigma = \left \{ 1.89 \left [\frac{h_1}{\rm (GeV)^3} \right ]
      +7.66\left [ \frac{h_3}{\rm (GeV)^3} \right ] \right \}
        {\rm pb},
\end{equation}
for $m_c=1.6$GeV and $\alpha_s(m_c)=0.24$. If we take numerical value
for $R_{cc}(0)$ estimated in \cite{K3} and use Eq.(12), one can get
numerical value for the contribution from $h_3$. But the contribution
from $h_1$ is unknown. If we take $h_1=h_3$, the cross section is
$0.23$pb. This value is larger than that estimated in \cite{BLO}.
The reason is that in \cite{BLO} the contribution of $h_1$ is not
present. We also calculate the angular distribution
and energy distribution, which are given in Fig.3 and Fig.4 respectively,
where $\theta_k$ is the angle between the moving direction of the $e^+$-beam
in CMS and that of $H_{QQ}$ and $x_k =2k^0/\sqrt{s}$.
From these figures one can see that the effect of $h_1$ is significant, especially
in the angular distribution, if $h_1$ is not much smaller than $h_3$.
\par
Our numerical results should be understood as rough estimations, because
the exact values of $h_1$ and $h_3$ are unknown. With the definitions
of these parameters in Eq.(9) one can use nonperturbative methods
to study them, or they can be extracted from experimental results
because they are universal. Once they are known, accurate results
can be given with our results.

%%%%%%%%%%%%%%%%%%%%%%%%%% Insert Fig 3 %%%%%%%%%%%%%%%%%%%%%%%%%%%%%%%%%%%%%
\begin{figure}[htbp!]
\center{\includegraphics[width=8cm]{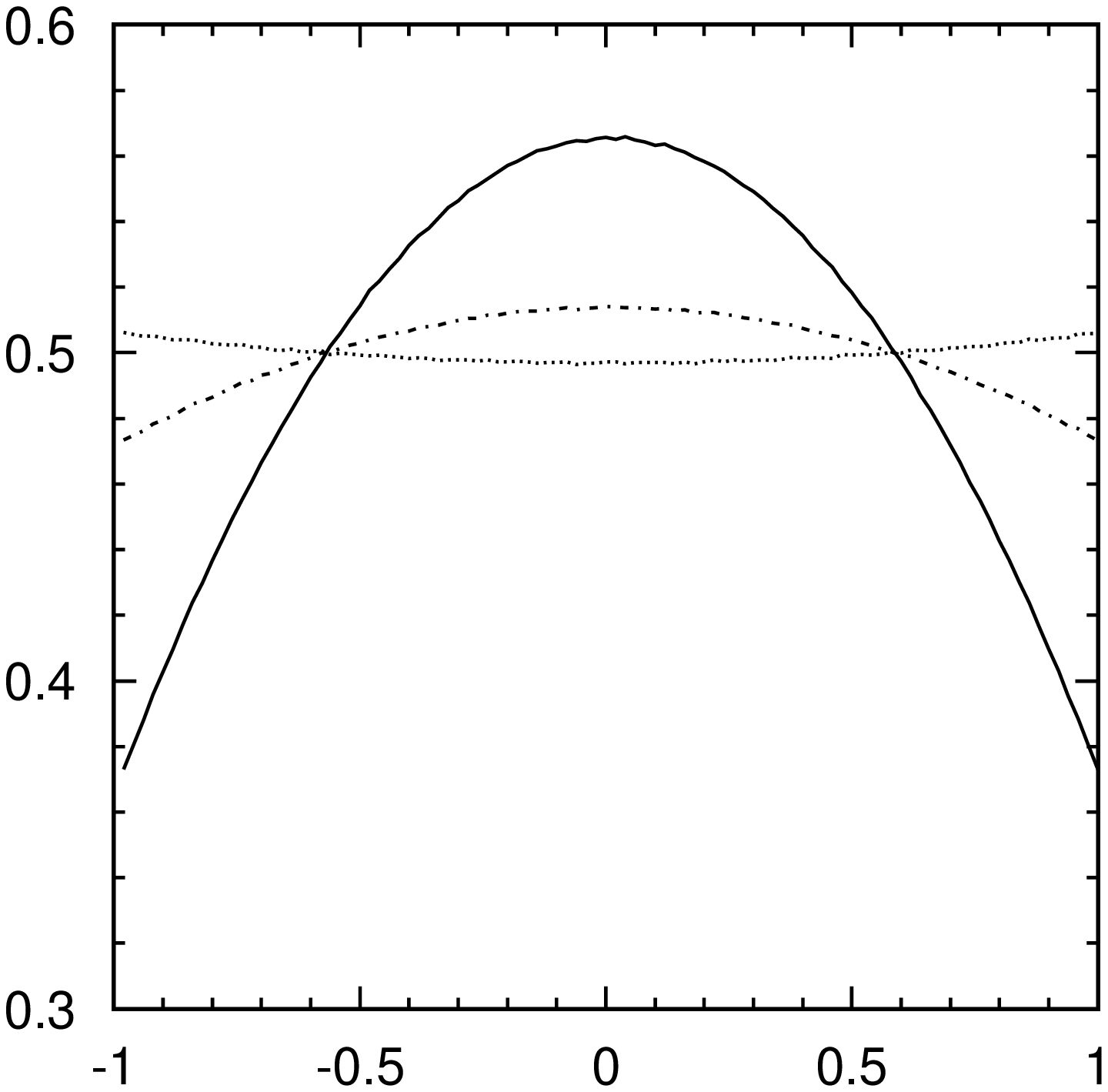}}
% \centerline{\psfig{figure=Fig3.eps,width=8cm,height=8cm}}
\caption{Results for the distribution
$\frac{1}{\sigma}\frac{d\sigma}{dcos\theta_k}$.
The solid line is for $h_3=0$, the dotted is for $h_1 =0$  and the
dash-dotted is for $h_1=h_3$.}
\end{figure}

%%%%%%%%%%%%%%%%%%%%%%%%%% Insert Fig 4 %%%%%%%%%%%%%%%%%%%%%%%%%%%%%%%%%%%%%
\begin{figure}[htbp!]
\center{\includegraphics[width=8cm]{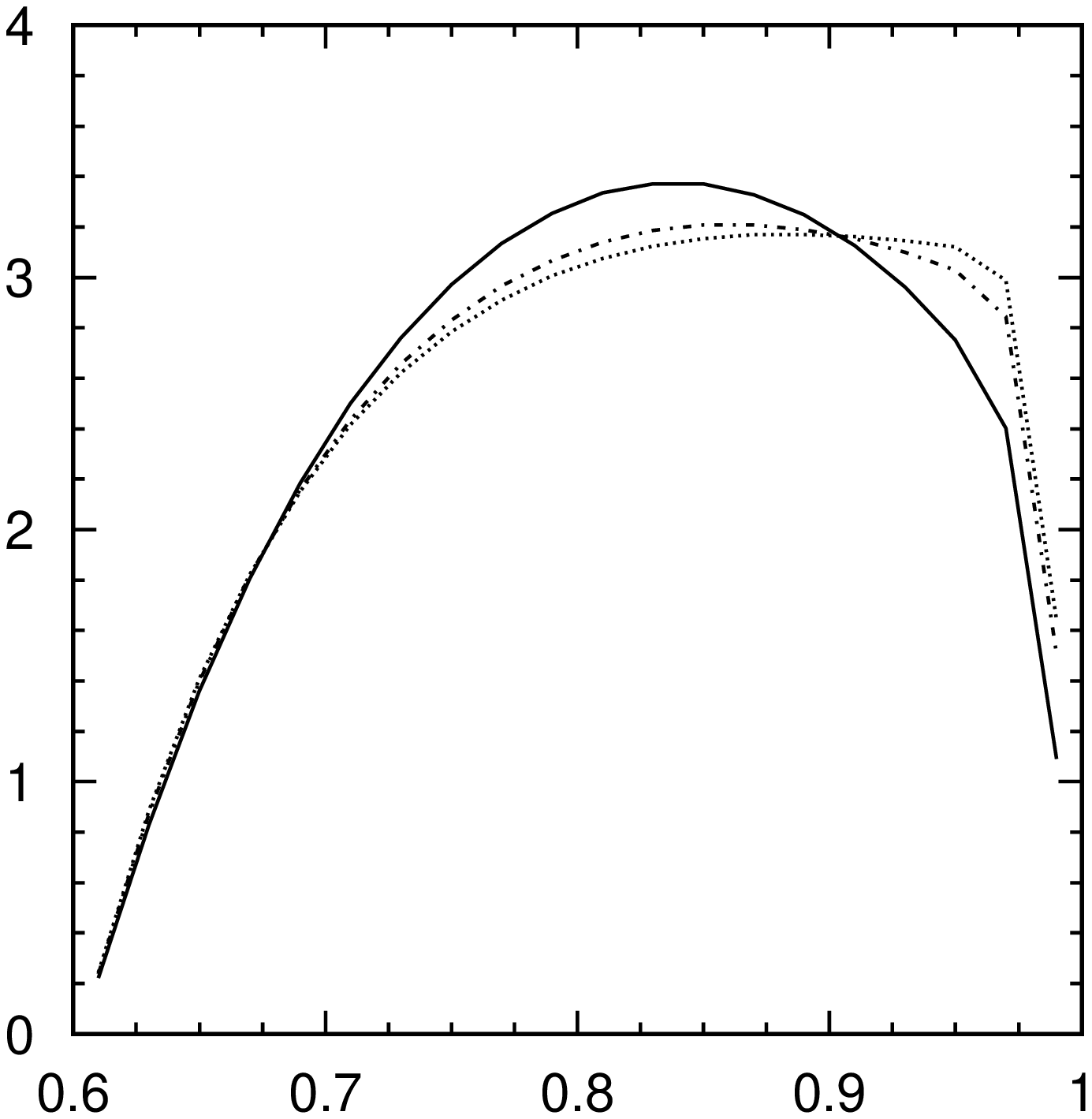}}
\caption{Results for $\frac{1}{\sigma}\frac{d\sigma}{dx_k}$.
The solid line is for $h_3=0$, the dotted is for $h_1 =0$  and the
dash-dotted is for $h_1=h_3$.}
\end{figure}

%%%%%%%%%%%%%%%%%%%%%%%%%%%%%%%%%%%%%%%%%%%%%%%%%%%%%%%%%%%%%%%%%%%%%%%%%%%%
%%%%%%%%%%%%%%%%%%%%%%%        Section 3           %%%%%%%%%%%%%%%%%%%%%%%%%
%%%%%%%%%%%%%%%%%%%%%%%%%%%%%%%%%%%%%%%%%%%%%%%%%%%%%%%%%%%%%%%%%%%%%%%%%%%%

\par\vskip20pt
\noindent
{\bf\large 3. Heavy Quark Fragmentation}
\par\vskip20pt
In this section we will use the factorization approach to calculate
the fragmentation function of a heavy quark $Q$ into a doubly heavy
baryon $H_{QQ}$. We will calculate the function by starting from
its definition\cite{CS}.
To give the definitions for a fragmentation function it is convenient
to work in the light-cone coordinate system. In this coordinate system
a 4-vector $p$ is expressed as $p^{\mu}=(p^+,p^-,{\bf p_T})$, with
$p^+=(p^0+p^3)/\sqrt{2},\ p^-=(p^0-p^3)/\sqrt{2}$.
Introducing a vector  $n$ with
$n^{\mu}=(0,1,{\bf 0_T})$, the fragmentation function
can be defined in the light cone gauge $n\cdot G(x)= 0$ as\cite{CS}:
\begin{eqnarray}
   D_{H_{QQ}/Q}(z) &=&  \frac {z}{ 4\pi} \int dx^- e^{ -ik^+x^-/z}
    \frac{1}{3}{\rm Tr_{color}}\frac{1}{2}{\rm Tr_{Dirac}}
       \nonumber \\
     && \cdot  \{
    n\cdot \gamma  \langle 0\vert Q(0)
    a_H^\dagger (k^+, {\bf 0_T}) a(k^+,  {\bf 0_T})
   \bar Q (0, x^-,{\bf 0_T}) \vert 0 \rangle \},
\end{eqnarray}
where $G_{\mu}(x)=G^a_{\mu}(x) T^a$, $G_{\mu}^a (x)$ is the
gluon field and the $T ^a (a=1,\dots, 8)$ are the color matrices.
The summation over the spin of $H_{QQ}$ is implied. In other gauges
gauge links must be supplied to make the definition gauge invariant.
The function
 $ D_{H/QQ}(z)$ is interpreted as the probability
 of a quark $Q$ with momentum $p$ to decay into the hadron $H$
with momentum component $k^+=z p^+$.
The function is invariant under a Lorentz boost along
the $z$-direction. Hence we can calculate the function in the
rest frame of $H_{QQ}$. It should be noted that the above definition
is given with relativistic normalization of states. Starting definitions
of fragmentation functions, various fragmentation functions for quarkonia
have been calculated\cite{MaF,BL,JL}. In the case of doubly heavy baryon
the calculation is similar.
\par
%%%%%%%%%%%%%%%% Inset Fig. 3 here %%%%%%%%%%%%%%%%%%%%%%%%%%%%%%
\begin{figure}[hbt]
\centering
\includegraphics[width=4cm]{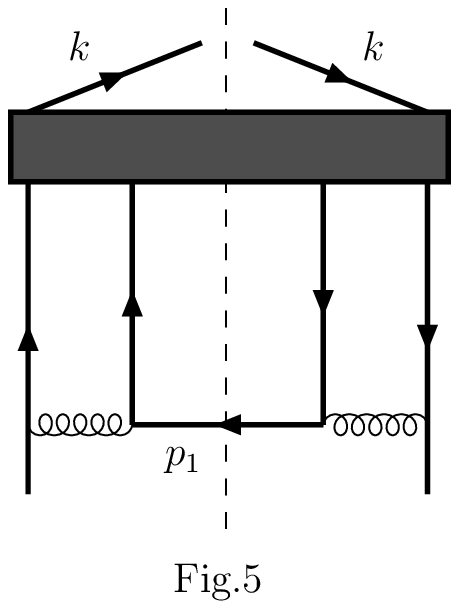}
\caption{Graphic representation for fragmentation function.
}
\label{Feynman-dg2}
\end{figure}
%%%%%%%%%%%%%%%%%%%%%%%%%%%%%%%%%%%%%%%%%%%%%%%%%%%%%%%%%%%%%%%%%%%%%%%%%%
\par
At tree-level, the fragmentation can be understood as the following:
The heavy quark generates a heavy quark pair through
exchange of a hard gluon, the two heavy quarks will be combined
with other partons generated nonperturbatively into the baryon. This
contribution is illustrated in Fig.5.  It is straightforward to
write down the contribution
to the function from its definition:
\begin{eqnarray}
D_{H_{QQ}/Q}(z) &=&  \frac {z g_s^4}{ 24\pi} \int dx^- e^{ -ip^+x^-/z}
   \int \frac{d^4p_1}{(2\pi)^4} 2\pi \delta (p_1^2-m_Q^2)
      \nonumber\\
 &&\cdot \int d^4 x_1 d^4 x_2 d^4 y_1 d^4 y_2 e^{ip_1\cdot x_1} \langle 0\vert
   \bar Q_i(x_1) \bar Q_j (x) [a^\dagger a] Q_k (y_1) Q_l(y_2) \vert 0\rangle
   \nonumber\\
 && \cdot  (\gamma^{\mu_1} T^a (\gamma\cdot p_1 -m_Q) \gamma^{\mu_2} T^b)_{il}
   (\gamma^{\nu_1} T^a S(x_2,x) \gamma\cdot n S(0,y_1) \gamma^{\nu_2} T^b)_{jk}
   \nonumber\\
 && \cdot D_{\mu_1\nu_1}(x_1,x_2) \cdot D_{\mu_2\nu_2}(y_2, y_1),
\end{eqnarray}
where $x^\mu =(0,x^-,{\bf 0_T})$. $iS(x,y)$ is the quark propagator of $Q$,
while $-iD_{\mu\nu}(x,y)$ is the gluon propagator in the
light cone gauge.
By using the results for the matrix element obtained before,
we obtain:
\begin{eqnarray}
D_{H_{QQ}}(z,\mu) &=& \alpha_s^2(\mu ) \frac{z(1-z)^2}{(2-z)^6} \cdot \{
        \frac{8 h_1}{27m_Q^3} (3z^4-8z^3+8z^2+48)
        \nonumber\\
    &&    +\frac{16 h_3}{9m_Q^3} ( 5z^4-32z^3+72z^2-32 z+16) \}.
\end{eqnarray}
The calculation can be done in the rest frame of $H_{QQ}$ straightforwardly,
where we have to convert  the matrix element in Eq.(20) through the factor
$4m_Q$ into that with nonrelativistic normalization. The term with $h_3$
is also calculated in \cite{FLSW}, it is the same as ours.
\par
With these results, one can estimate the production rate at a hadron collider
like Tevatron and LHC. The rate can be estimated as:
\begin{equation}
  \sigma_{H_{QQ}} (p_t) \approx \sigma_{Q} (p_t) \cdot M_Q^{(1)}(p_t),
\end{equation}
where $\sigma_{H_{QQ}} (p_t)$ and $ \sigma_{Q} (p_t)$
is the cross section for inclusive production
of $H_{QQ}$ and $Q$ with transverse momentum larger than $p_t$, respectively.
$M_Q^{(1)}(p_t)$ is the first moment of the fragmentation
function at the energy scale $\mu =p_t$. To avoid large logarithm like $\ln {p_t/m_Q}$
in our perturbative result of $D_{H_{QQ}}$, one can use renormalization group
method to sum these large log terms. But, if in the summation one neglects the gluon
fragmentation which is at least at order of $\alpha_s^3$ and uses
one loop result for anomal dimensions,
then the first moment does not change with the scale $\mu$. We use $\alpha_s(\mu)$
at $\mu =m_c$ for calculating the first moment.
$\sigma_{Q} (p_t)$  is calculated at tree-level
by taking two partonic processes $q\bar q\to Q\bar Q$
and $gg\to Q\bar Q$. Taking $p_t=40$GeV, $m_c=1.6$GeV and $\alpha_s(m_c)=0.24$
we obtain
for LHC and Tevatron:
%\begin{eqnarray}
%\sigma_{\Xi^+_{cc}} &\approx & \left\{0.0039 \left [ \frac{h_1}{\rm (GeV)^3} \right ]
%   +0.0080 \left [ \frac{h_3}{\rm (GeV)^3} \right ]\right\}{\rm mb}, \ \ \ \
%   {\rm for\ Tevatron},
% \nonumber\\
%\sigma_{\Xi^+_{cc}} &\approx & \left\{0.046 \left [ \frac{h_1}{\rm (GeV)^3} \right ]
%   +0.095 \left [ \frac{h_3}{\rm (GeV)^3} \right ]\right\}{\rm mb}, \ \ \ \
%   {\rm for\ LHC}.
%\end{eqnarray}
\begin{eqnarray}
\sigma_{\Xi^+_{cc}}&\simeq &\Big\{ 0.0014\Big[\frac{h_1}{(GeV)^3}\Big]
+0.0029\Big[\frac{h_3}{(GeV)^3}\Big]\Big\},~~~~{\rm for~Tevatron}, \nonumber \\
\sigma_{\Xi_{cc}^+}&\simeq & \Big\{0.020 \Big[\frac{h_1}{(GeV)^3}\Big]
+0.042 \Big[\frac{h_3}{(GeV)^3}\Big]\Big\},~~~~{\rm for~LHC}.
\end{eqnarray}
If we take the same value of $h_3$ as in the last section and $h_1=h_3$,
we obtain $\sigma_{\Xi^+_{cc}}=0.0018$mb for LHC and
$\sigma_{\Xi^+_{cc}}=0.00013$mb for Tevatron, respectively.
With the planed luminosity $100({\rm fb})^{-1}$ per
year of LHC
there will $1.8=\times 10^{11}$ $\Xi^+_{cc}$'s produced at LHC. Our numerical
results are roughly 10 times larger than those estimated in \cite{K3}. Beside
the extra contribution from $h_1$, one main reason
for this is that our estimations are sensitive to values of parameters. If we take
$m_c=1.7$GeV and $\alpha_s =0.2$ as taken in \cite{K3}, our numerical
results will become 10 times smaller. It is interesting to note that
our numerical results roughly remain the same if we take $p_t=10$GeV. However,
our estimation for $p_t=10$GeV or smaller $p_t$ can not be reliable, because
contributions from fragmentation are dominant at high $p_t$ and other contributions
are significant at low $p_t$. Including all contributions one is able
to show that contributions from fragmentation become dominant
for $p_t\sim 25-30$GeV\cite{K3}.
\par\vskip20pt
\noindent
{\bf\large 4. Summary}
\par\vskip20pt
We have studied inclusive production of doubly heavy baryon $H_{QQ}$
at a $e^+e^-$
collider and through fragmentation of a heavy quark, in which
a factorization was performed to factorize perturbative- and
nonperturbative effects. In our approach the production can be understood
as a two-step process, in which a $QQ$ pair is produced first and then
the pair is transformed into $H_{QQ}$ nonperturbatively. The production
of the $QQ$ pair can be studied with perturbative QCD because the large mass
$m_Q$. With the large mass $m_Q$ a heavy quark $Q$ moves with a small
velocity $v_Q$ in $H_{QQ}$ in its rest frame. This suggests that one can
use NRQCD to describe the transformation, where a systematic expansion
in $v_Q$ can be performed. At the leading order we find
that $H_{QQ}$ can be formed from two states of the $QQ$ pair,
one state is with the pair in $^3S_1$ state and in color ${\bf \bar 3}$,
another is with the pair in $^1S_0$ state and in color ${\bf 6}$.
The transformation from these two states are described by two matrix
elements $h_1$ and $h_3$, defined with NRQCD. A power counting
in $v_Q$ for these two matrix elements is given.
Our results are different than those in previous approaches,
where $H_{QQ}$ is formed only from the state of
the pair in $^3S_1$ state and in color ${\bf \bar 3}$.
Perturbative coefficients in the contributions of these two states
to the production at $e^+e^-$ colliders and to the production
through heavy quark fragmentation are calculated at tree-level.
\par
Numerical results are given for $\Xi^+_{cc}$-production at
B-factories and for $\Xi^+_{cc}$-production through fragmentation
at LHC and Tevatron, in which we relate the parameter $h_3$
to a wave function of the $cc$ pair, which has been studied
and results are given for different ratios $h_1/h_3$. We find
that the contribution of $h_1$, i.e., of the state of the $cc$ pair
in $^1S_0$ state and in color ${\bf 6}$, is significant, if
$h_1$ is not much smaller than $h_3$. It should be noted
that detailed values of the two matrix elements are unknown,
our numerical results should be taken as rough estimations.
The two matrix elements can be studied by nonperturbative
methods, or extracted from experiment. If their values are known,
detailed predictions can be made.
\par\vskip20pt\noindent
{\bf Acknowledgements}
\par
This work is supported by National Nature
Science Foundation of P. R. China.

\par\vfil\eject
\noindent
{\bf\large Appendix}
\par\vskip20pt
The functions in the differential cross section are:
\begin{eqnarray}
{\cal A}_1^{(1)}&=&-\frac{32m_Q^6\Big[4(1-x_3-x_4)+(x_3^2+x_4^2)+
x_3 x_4(4-x_3-x_4)\Big]^2}{\hat{s}^3(1-x_3)^2(1-x_4)^2}+
\frac{16m_Q^4}{\hat{s}^2(1-x_3)^2(1-x_4)^2}\Big\{32\nonumber \\
&&-88(x_3+x_4)+78(x_3^2+x_4^2)-16(x_3^3+x_4^3)-16(x_3^4+x_4^4)+12(x_3^5+x_4^5)
-3(x_3^6+x_4^6)\nonumber\\
&&+x_3 x_4\Big[228-192(x_3+x_4)+44(x_3^2+x_4^2)+20(x_3^3+x_4^3)
-12(x_3^4+x_4^4)+2(x_3^5+x_4^5)\nonumber\\
&&+x_3 x_4[168-56(x_3+x_4)+(x_3^2+x_4^2)+
2(x_3^3+x_4^3)+4x_3 x_47-x_3-x_4)]\Big]\Big\}\nonumber\\
&&+\frac{2m_Q^2}{\hat{s}(1-x_3)(1-x_4)}\Big\{-64
+160(x_3+x_4)-116(x_3^2+x_4^2)+24(x_3^3+x_4^3)+7(x_3^4+x_4^4)\nonumber\\
&&-3(x_3^5+x_4^5)+x_3x_4\Big[-360+216 (x_3+x_4)-24(x_3^2+x_4^2)
-11(x_3^3+x_4^3)+3(x_3^4+x_4^4)\nonumber\\
&&-2x_3x_4[55-7(x_3+x_4)+2(x_3^2+x_4^2)-x_3x_4]\Big]\Big\}
+(x_3-x_4)^2\Big\{-6
+10(x_3+x_4)\nonumber\\
&&-5(x_3^2+x_4^2)+(x_3^3+x_4^3)-x_3x_4(8-x_3-x_4)\Big\},\nonumber\\
{\cal A}_2^{(1)}&=&-\frac{16m_Q^4\Big[2-2x_3+
x_3^2-x_4^2\Big]^2}{\hat{s}^2(1-x_3)^2}+\frac{8m_Q^2}{\hat{s}(1-x_3)}
\Big\{6-x_3(10-9x_3+3x_3^2)-x_4(4+3x_4\nonumber\\
&&-2x_4^2)+x_3x_4(2+x_4)\Big\}
-x_3^2(1-x_3)-x_4^2(5-3x_4+x_4^2)-x_3x_4\Big[-6+x_4-2x_4^2\nonumber\\
&&+x_3(3+x_4)\Big],\nonumber\\
{\cal A}_3^{(1)}&=&2\Big\{8(x_3+x_4-1)-3(x_3^2+x_4^2)
+x_3x_4[(x_3-x_4)^2-2]\Big\}
+\frac{16m_Q^2}{\hat{s}(1-x_3)(1-x_4)}\Big\{
\frac{2m_Q^2}{\hat{s}}\Big[\nonumber\\
&&4(x_3+x_4-1)+(x_4^4+x_3^4)-2(x_3^3+x_4^3)
+2x_3x_4(x_3+x_4-2-x_3x_4)\Big]+6-10(x_3\nonumber\\
&&+x_4)+4(x_3^2+x_4^2)+2(x_3^3+x_4^3)-(x_3^4+x_4^4)+2x_3x_4[7-3(x_3+x_4)+x_3x_4
]\Big\},\nonumber\\
{\cal A}_4^{(1)}&=&2(1-x_3)\Big[x_3^3+x_4(8-8x_4+3x_4^2)-x_3x_4(3x_3+x_4)\Big]
+\frac{16m_Q^2}{\hat{s}(1-x_3)(1-x_4)}\Big\{\nonumber\\
&&\frac{2m_Q^2}{\hat{s}(1-x_3)}\Big[x_3(2-2x_3+x_3^2)^2
+x_4(4-4x_4+2x_4^3-x_4^4)-x_3x_4[12-x_3(12-6x_3\nonumber\\
&&+x_3^2)-x_4(4-2x_4+x_4^2)+2x_3x_4]\Big]
-x_3[6-10x_3+9x_3^2-3x_3^3]-x_4(6-10x_4\nonumber\\
&&+4x_4^2+2x_4^3-x_4^4)+
x_3x_4\Big[20-x_3(16-7x_3+2x_3^2)+x_4(2x_4^2+x_4-15)\nonumber\\
&&+x_3x_4(7-x-4)\Big]\Big\}.
\end{eqnarray}

\begin{eqnarray}
{\cal A}_1^{(3)}&=&
\frac{2m_Q^4}{\hat{s}^2(1 - x_3)^2(1 - x_4)^2}\Big\{
\frac{2 m_Q^2}{\hat{s}} \Big[-64+
 144(x_3+x_4)-100 (x_3^2+x_4^2)+20(x_3^4+x_4^4) \nonumber \\
&&-3(x_3^6+x_4^6)+x_3 x_4\Big(-328
+2[128(x_3+x_4)-28(x_3^2+x_4^2)-8(x_3^3+x_4^3)\nonumber\\
&&+3(x_3^4+x_4^4)])-x_3x_4[216+x_3^2+x_4^2-64(x_3+x_4)+20x_3 x_4]\Big)\Big]+32
-104(x_3+x_4)\nonumber \\
&&+118(x_3^2+x_4^2)-50(x_3^3+x_4^3)-28(x_3^4+x_4^4)+46(x_3^5+x_4^5)
-17(x_3^6+x_4^6)+x_3 x_4\Big[228\nonumber \\
&&-158(x_3+x_4)+44(x_3^2+x_4^2)+18(x_3^3+x_4^3)
-36(x_3^4+x_4^4)+14(x_3^5+x_4^5)-x_3 x_4[-32\nonumber \\
&&+8(x_3+x_4)-23(x_3^2+x_4^2)+4(x_3^3+x_4^3)+2x_3 x_4(5(x_3+x_4)-6)]\Big]\Big\}
+(x_3-x_4)^2\Big[\nonumber \\
&&(x_3^3+x_4^3)-3(x_3^2+x_4^2)+4(x_3+x_4)+3x_3 x_4[x_3+x_4-2]-2\Big]
-\frac{m_Q^2}{\hat{s}(1 - x_3)(1 - x_4)}\Big\{\nonumber \\
&&12 (x_3^2+x_4^2)-37(x_3^3+x_4^3)+42(x_3^4+x_4^4)-17(x_3^5+x_4^5)
+2(x_3^6+x_4^6)+x_3 x_4\Big[7x_3^4\nonumber \\
&&+7x_4^4-36(x_3^3+x_4^3)+34(x_3^2+x_4^2)
-3(x_3+x_4)+x_3x_4[4(x_3^2+x_4^2)+5(x_3+x_4)\nonumber\\
&&+6x_3x_4-32]\Big]\Big\}, \nonumber \\
{\cal A}_2^{(3)}&=&\frac{m_Q^2}{\hat{s} (1-x_3)}\Big\{
\frac{2m_Q^2}{\hat{s}(x_3-1)}\Big[12+3 x_3^4-16x_3^3
+36 x_3^2-32x_3+3x_4^4-4x_4^3-8x_4\nonumber \\
&&+2x_3 x_4[2x_3^2+x_3(x_4-10)
+2(6-2x_4+x_4^2)\Big]+\frac{1}{1-x_4}\Big[4-2x_3(x_3^3-x_3^2-3x_3+6)\nonumber\\
&&-x_4(12-16x_4+3x_4^2+9x_4^3-6x_4^4)+x_3 x_4\Big(
14+x_3(x_3-5)-x_4(10-5x_4+x_4^2)\nonumber \\
&&+x_3 x_4(5+x_3-6x_4)\Big)\Big]\Big\}+(x_3-x_4)^2(x_3+x_4-1), \nonumber \\
{\cal A}_3^{(3)}&=&2(x_3-x_4)^2 (x_3+x_4-1)+
\frac{m_Q^2}{\hat{S}(1-x_3)(1-x_4)}\Big\{
\frac{4m_Q^2}{\hat{s}}\Big[5 (x_3^4+x_4^4)-16(x_3^3+x_4^3)\nonumber \\
&&+22(x_3^2+x_4^2)-12(x_3+x_4)-4+2 x_3 x_4[20-12(x_3+x_4)
+3 (x_3^2+x_4^2)+x_3 x_4]\Big]\nonumber \\
&&+2(x_3^5+x_4^5)-3(x_3^4+x_4^4)+9(x_3^3+x_4^3)-20(x_3^2+x_4^2)+
16(x_3+x_4)-x_3x_4\Big[72\nonumber \\
&&+5(x_3^3+x_4^3)+4(x_3^2+x_4^2)-55(x_3+x_4)
+x_3 x_4[34-3(x_3+x_4)]\Big]\Big\},\nonumber \\
{\cal A}_4^{(3)}&=&-2(x_3-x_4)^2\Big[x_3^2-(1-x_4) x_4-x_3(1-2x_4)\Big]
+\frac{m_Q^2}{\hat{s}(1-x_3)(1-x_4)}\Big\{
\frac{4m_Q^2}{\hat{s}(1-x_3)}\Big[(12\nonumber \\
&& -32 x_3+36 x_3^2-16 x_3^3+3x_3^4) x_3
+x_4 (4+12x_4-22x_4^2+16x-4^3-5x_4^4)\nonumber\\
&&+x_3 x_4\Big(-12+
x_3[22-18x_3-x_3^2+2x_3^3]+x_4[-44+42x_4-15x_4^2+2x_4^3]\nonumber\\
&&+2x_3 x_4[16-4x_3+x_3^2-7x_4+x_4^2]
\Big)\Big]+
4x_3[-2+6x_3-3x_3^2- x_3^3+x_3^4]
+x_4^2[-16\nonumber\\
&&+20x_4-9x_4^2+3x_4^3-2x_4^4]
+x_3 x_4\Big(8+x_3[-8+x_3+x_3^2-2x_3^3]
+x_4 [40-49x_4+22x_4^2\nonumber\\
&&-7x_4^3]+x_3 x_4 [-35+9x_3 x_4-3x_3(2-x_3)+x_4(24-x_4)\Big)\Big\}.
\end{eqnarray}

%%%%%%%%%%%%%%%%%%%%%%%%%%%%%%%%%%%%%%%%%%%%%%%%%%%%%%%%%%%%%%%%%%%%%%%%%%%%
%%%%%%%%%%%%%%%%%%%%%%%%%%%%%%%%%%%%%%%%%%%%%%%%%%%%%%%%%%%%%%%%%%%%%%%%%%%%
%%%%%%%%%%%%%%%%%%%%%%%%%%%%%%%%%%%%%%%%%%%%%%%%%%%%%%%%%%%%%%%%%%%%%%%%%%%%
\vfil\eject

\end{document}